\documentstyle[11pt]{article}
 
\def\pa{\partial}
\def\k{\kappa} 
\def\g{\gamma} 
\def\a{\alpha} 
\def\b{\beta} 
\def\d{\delta} 
\def\e{\epsilon} 
 
\def\k{\kappa}
 \def\L{\Lambda}
\def\m{\mu} 
\def\n{\nu}

 \def\S{\Sigma}

\def\o{\omega} 

\def\mn{{\mu\nu}}
     
\begin{document}


\begin{center}
{\large\bf Physico--Mathematical Interactions:\\ 
 The Chern--Simons Story}

S. Deser\\

Department of Physics\\
Brandeis University\\
Waltham, MA 02454, USA
\end{center}

\begin{quotation}
The essential role played by Chern--Simons terms in
a variety of physical models provides yet another
illustration of the unexpected but profound interactions
between the two disciplines.
\end{quotation}

Ludwig Faddeev is one of the rare few who are regarded
as mathematicians by mathematicians and as physicists by
physicists (more commonly, it is vice versa!).  He has
contributed to the synthesis between our two disciplines
in many domains; I had the pleasure of working with him
on problems in general relativity and of many discussion
over the years.

I propose to illustrate this synthesis through a 
particular set of 
examples,  Chern--Simons (CS) 
``effects" in physics.  This
should both reflect the interplay of the two disciplines,
as well as the uncanny way mathematical constructs
become incorporated into physics (and sometimes even
require the physicist to be a little precise). 
I must of necessity be succinct here, and refer 
(also compactly) to the literature for details. I
shall not have the space here to illustrate the
``backreaction", how such borrowing by physics in turn
stimulates new mathematics; ``CS mathematics"
distinctly picked up after the advent of CS physics.

To do justice to the full web of interconnections
involving (CS) terms \cite{001} in
physics would require one of those complicated tree
(or loop) diagrams. I will have to omit 
entirely any discussion of some
of the principal ones, for example
the relation of CS to ~a) conformal
field theory \cite{002} (descending from 3 to 2
dimension in particular) ~b) anomalies \cite{003}, 
via its Pontryagin $F\wedge F$ ancestor (ascending 
from 3 to 4), ~c) to integrable systems \cite{004},
and ~d) in the currently hot AdS ``$M$-theory" context.
Instead, I will stick to some more concrete applications
in which I have been involved.

The first sighting of  CS in physics may 
have been in 1978, when D=11 supergravity (now back 
in a central role after two decades, thanks to $M$ 
theory) was constructed.  
It arose there as a strange but unavoidable
term needed for consistency of the theory, by preserving
its local supersymmetry, then rapidly invaded lower
dimensional, 4$<$D$<$11, models  \cite{005}.  
That a metric-independent,
``topological",  term (as physicists sometimes call them)
should come to the rescue of a gravitational model is
the first example of its uncanny properties!  The theory
necessarily contains a 3-form potential $A$, and it was
found that there has to be an addition
$_{11}I_{CS} [A] = \k \int A \wedge F \wedge F$,
$F \equiv dA$ to the usual $F^2$ kinetic term  
in the action. The Einstein gravitational
constant $\k$ appears here, but not (of course) the
metric.  A smaller paradox is that despite
appearances, $I_{CS}$ is both parity and $T$ even.
From a physical point of view, this term generates a
cubic self-interaction of the form field that is in fact
essential in constructing its 
supersymmetry-preserving invariants \cite{006}.  These 
invariants are important as they can serve both 
as a check of M-theory
currently thought to incorporate the D=11 theory as
a limiting case and as counter terms in higher 
loop corrections to this maximal supergravity
itself \cite{006,007}.  
This first physics appearance of CS
passed relatively unnoticed for several reasons, not
least the cubic nature of $_{11}I_{CS}[A]$, 
so that it did not directly affect the kinematics.
Soon afterwards, and with no apparent connection
to the above, the possibility and interest of  
introducing the 1-form CS term in spacetime
dimensions D=3 was suggested by several authors
\cite{008,009}.  This time the context was
more auspicious both because D=3 is closer to D=4 and
because physics in this planar world may even
have observable consequences, in condensed matter
settings as well as in high temperature limits of
our D=4 world.  Most of all, the interest was due to
the fact that CS is here quadratic (and $P, \; T$
violating), 
$~_{3}I_{CS}[A]  = \m \int A \wedge F$ and hence
can affect free-field (Maxwell) electromagnetism, 
and indeed
lead to a finite-range but still gauge-invariant model.
In its nonabelian incarnation, where $A $ is a
Lie algebra-valued {1-form}, the term has the 
remarkable property that its numerical coefficient
must be quantized for the quantum theory to be
well-defined \cite{009}.  This idea, coming
entirely from homotopy analysis, was of course a
revelation to physicists on how  
{\it a priori} arbitrary parameters could
in fact be restricted in their possible values
(and hence had better also be renormalized by
integer amounts only).

Before we consider some of the novel
consequences of CS in this D=3 context, we first
mention a quite different direction that 
gave rise to an enormous literature on so-called 
topological quantum field theories, including
D=3 gravity, as we shall see.  For the moment,
we take the
geometry to be Minkowskian $R^1\times
R^2$, to avoid global and topological complications.
Then the Euler--Lagrange equations of purely CS
actions simply become $^*\! F^\m = 0$, in the
absence of sources or $^*\! F^\m = j^\m$ when 
charged currents are present ($^*\! F$ is the dual
field strength, a 1--form).  Thus the field is locally pure
gauge wherever there are no sources; to find the general
global solution with the properties 
that the field strength is equal to the current and 
vanishes elsewhere is then an
interesting exercise.  This is even more so in the 
nonabelian case where the abelian part is
supplemented  by the famous 
$\frac{1}{3} \: tr \int (A\wedge A\wedge A)$
addition to yield the same (but now nonabelian) 
Euler--Lagrange equation $^*\!F =0$.  Now in 
D=3, general relativity has a very similar property:
spacetime is flat in the absence of source, since
Einstein ($G$) and Riemann ($R$)  tensors are
equivalent, obeying the double-duality identity
$G \equiv ^*\!\!R^*$.  Hence the
Einstein equations $G = 0$ imply local
flatness.  Classification of such Minkowski signature
locally flat, (or more generally locally constant 
curvature if there is also a cosmological constant
so that $G + \L g=0$), spacetimes \cite{010} has also 
become a physical industry of its own (we cannot
even start to cite this literature). 
Here the physics involves global matching of flat
patches at particle trajectories where the sources
$T_\mn$ (and therefore curvature) do not vanish. This
``zero field-strength" field equation in source-free
gravitational regions is of
course very reminiscent of the above 
Yang--Mills CS
story and indeed there is a CS form of 
(except for some fine print) Einstein
gravity \cite{011}.  This insight has led to
another large topic of its own ever since, namely
the uses of the ``antigeometrical" CS as geometry!
[For a review of these formulations and their 
properties as well as references, see \cite{new11}].
There is also a direct mathematical connection,
namely that between the Riemann--Hilbert problem
and D=3 gravity coupled to several moving 
particles \cite{012}.

This is by no means the end of the gravity-CS 
interaction.  There also exists in D=3 a genuine 
{\it gravitational} CS term 
\begin{equation}
I_{CS} \equiv - \textstyle{\frac{1}{4}} \int
d^3 x \: tr \e^{\m\n\a} 
[R_{\m\n} \o_\a + \textstyle{\frac{2}{3}}
\o_\m \o_\n \o_\a ]\; ,
\end{equation}
where the traces over the local Lorentz indices
and $\o_\m$ is the spin connection. Its variation 
yields the Cotton tensor 
$\sqrt{g} \, C^{\m\n} \equiv \e^{\m\a\b}
D_\a (R^\n_\b - \frac{1}{4} \d^\n_\b R)$
(invented incidentally by a French mathematician 
whose brother was a physicist).  This
tensor is (despite appearances) symmetric,
conserved and traceless; it is in fact the 
conformal curvature tensor in D=3, where the
usual Weyl tensor vanishes.  A pure CS action,
whose sources must necessarily be traceless
(and so cannot be particles or photons) is
therefore not so interesting; it has exterior
solutions to the third order equation $C_{\m\n}
= 0$ that are conformally flat.  Far more 
interesting is topologically massive gravity
(TMG) \cite{009},  the sum of Einstein and 
gravitational CS actions.  This model is in a
sense the opposite of the topologically massive
electrodynamics to be discussed below: the
highest (third) derivative is in the CS part
here.  Despite the higher derivative content,
TMG is perfectly ghost-free
and consists of a massive helicity $\pm$2
excitation according to the relative sign of
the coefficients between the two terms; here 
$P$ and $T$ are each violated.  TMG
differs from the non-abelian TM vector case in
{\it not} requiring quantization of the CS
coefficient, despite seemingly similar arguments;
this is quite paradoxical at first sight. Indeed,
TMG still presents some other challenges; 
in particular
no one has yet found the ``Schwarzschild 
solution" for this nonlinear theory, although
there is an amusing anyonic structure in the 
linear theory's solutions \cite{new12}, with the
CS term ``twisting" the spin of source particle,
as for the vector case.  Other challenges include
deciding whether the theory is renormalizable or
not \cite{new13}; it might appear to be so on the 
basis of its higher derivative structure, but the 
latter does not quite ``shield" the conformal
contributions.  The issue can in principle be 
decided using homotopy arguments.  
The interest here is that this is the only unitary
but higher derivative theory of gravitation that
has a dimensional coupling constant and would be
the only finite quantum gravity model.
There are of
course (higher order in curvature) generalizations
of the Cotton tensor to other odd dimensions,
but they do not affect the propagators; there are 
generalizations to higher spin fields in
D=3 as well (for $s$=3, see \cite{new13}).

Returning to physical applications of the plain
abelian CS term, let me sketch a few of the 
reasons for their interest.  First, if we add the
usual Maxwell action to CS, the resulting 
topologically massive electrodynamics (TME)
represents a single local degree of freedom,
paradoxically endowed with a finite range but 
still gauge invariant.  
[This seems a very different way to get a finite
mass than the Higgs mechanism but things
are even more interesting; see \cite{013}].
As background, recall that
a pure Maxwell excitation in any dimension has
(D-2) local excitations, the transverse spatial
polarizations, while the gauge-broken (Proca
theory)  massive version has (D-1) of them.  Further, 
the latter theories represent excitations of unit
intrinsic angular momentum or spin.  In D=3,
however, it turns out that massless 
fields, including Maxwell,
are (unlike massive ones) entirely devoid of spin
\cite{014} (but neutrinos still can have 
fermi statistics).  The TME action 
inherits from pure Maxwell one local excitation;
the CS input is to provide mass and thereby
spin to this excitation.  Here the CS term
does break parity: the two degrees of the 
normal Proca theory are equivalent to a pair
of ``mirror" TME models. The 
TME field equations 
$dF + \m ^*\!F = 0$ are readily seen to imply
that the field strength obeys Klein--Gordon
propagation equations with $\m$
representing the finite range.  This mixing of
normal metric and topological terms is what
makes these models so different from the usual
even-dimensional ones.

Mathematically, we
have mentioned the role played by homotopy
in CS physics.  In fact there are several 
different roles, as we shall see.  One
is the cited
quantization of the CS coefficient in the
nonabelian theory: because the exponential
of the action, $e^{iI/\hbar}$ is the basic 
quantum mechanical object there, actions
must be invariant mod $2\pi$ under gauge
transformations.  Tracing the $\Pi_3/\Pi_1$
properties of CS under large gauge 
transformations shows it changes by a winding
number so that its coefficient is necessarily
integer; this is the dimensionless combination
$\m /g^2$ where $g^2$ is the 
(dimensional in D=3) self-coupling
constant.  Another effect is that of the 
topology of planar configuration space
-- this is related to
``anyons" or the loss of the standard 
spin-statistics relation in planar field theories
\cite{015}; it too can be represented in CS
language \cite{016}.

My final example is the most recent; it deals
with the definition and role of abelian 
CS in nontrivial topologies (the nonabelian
CS story is still in progress \cite{new16}).
The important issues already appear 
in cases as simple as
$S^1 \times \S^2$, finite-temperature
($\b = \frac{1}{kT}$ is the perimeter
of $S^1$) planar electrodynamics with
(necessarily quantized) magnetic flux
in the closed 2-manifold $\S^2$. 
For details and earlier literature, 
I must refer to \cite{018}.  It is 
known that the naive CS term 
$\int A \wedge F$ now
requires corrections to remain well-defined.
These corrections to CS, and its behavior
under large (not reducible to the 
identity) gauge transformations can in fact
be elucidated in two complementary ways,
(and different from the known cohomology
procedures cited in \cite{017}).  
The first uses a classic
result, the Chern--Weil theorem, which in
D=4 tells us 
(using transgression)
that for two different connections
$(A,\tilde{A})$ on a bundle,  
$F \wedge F - \hat{F} \wedge \hat{F} = d[(A-\hat{A})
\wedge (F+ \hat{F})]$.  This 
provides a correct definition of $I_{CS}$ on
non-trivial bundles and also tells us that,
unlike the simple-minded CS, it
changes under large gauge rotations as
the product of their ``winding number" 
and the magnetic flux so as to respect the
quantum action requirements mentioned
above.  The second way to reach the correct
definition is -- surprisingly -- to embed
the abelian model in a nonabelian
SU(2) where all D=3 bundles are trivial;
the fact that the homotopies of U(1)
and SU(2) are opposite ($\Pi_1$ of the
former and $\Pi_3$ of the latter fail to
vanish) is no obstacle.  [There
is a third, heuristic, way -- the one a desperate
physicist would use to ``guarantee" correctness
when all else fails \cite{017,018}, but I do not
discuss it here!]  I cite this seemingly pedantic 
formal discussion of CS definition
precisely because what is the
correct one in 
topologically nontrivial backgrounds has 
led to an immense and rather confused
physics literature; confused because based on
the naive $\int A \wedge F$, immense because it
concerns the physically important ``thermal" 
quantum electrodynamics where time is replaced by
temperature through periodic identification
of $t$, as we have mentioned.  Now the
CS miracle  here is that, whether or not there
is a ``primitive" CS term in the
original action (or indeed any action at all
for the electromagnetic field $A$), there
will arise an effective theory of $A$ if one 
integrates out the charge particles that
(necessarily) couple to it.  In particular,
if we have massive charged electrons
obeying the usual Dirac equation
$(D\!\!\!\!/ + m) \psi =0, \; D\!\!\!\!/ \equiv
\g (\pa + iA)$, then the (logarithm of the)
determinant of the Dirac operator is 
essentially the functional that defines the
effective action $I_{eff}[A]$.  Since a
fermion mass
term is parity (and $T$) violating in D=3,
there should naturally be CS terms in 
$I_{eff}$.  Now the route to this action
involves a careful process of first defining
the determinant, {\it e.g.}, by 
$\zeta$-function regularization.  This enables
one to expand in Seeley--deWitt coefficients,
and find the correct, automatically
gauge-invariant $I_{eff}[A]$.  In particular the
CS term always enters in a way that
preserves invariance namely as part of deeper
$\eta$-function structures.  It was neglecting or
omitting this necessary complication that gave
rise to paradoxes involving large gauge
transformations, that of course do 
{\it not} leave CS invariant.  Indeed, to a
physicist CS is basically the reminder
already in the abelian but globally non-trivial
space context that there is a further, discrete,
gauge invariant
variable besides the field strengths, namely
the so-called flat connection rather than CS
itself.

I have tried to 
give one short glimpse of how  a
theoretical physicist is often forced to
use -- and greatly benefits from -- 
{\it a priori} far-removed mathematical
constructs, a process that ultimately leads
to further advances in mathematics as well.

\noindent{\bf Acknowledgements}

I thank R. Jackiw and especially 
D. Seminara for useful comments, 
as well as all the other coauthors who have 
contributed to my
Chern--Simons education over many contexts and years.
This work was supported  by NSF grant PHY 93-15811.

\end{document}